\documentclass[%
 aip,
 jmp,%
 twocolumn,
 amsmath,amssymb,
 reprint,%
]{revtex4-1}

\usepackage[dvips]{color}
\usepackage{graphicx}
\usepackage{dcolumn}
\usepackage{bm}
\usepackage{ulem}

\begin{document}

\title{Delocalized-localized transition in a semiconductor two-dimensional honeycomb lattice} 

\author{G. De Simoni}
\affiliation{NEST, Istituto Nanoscienze-CNR and Scuola Normale Superiore, I-56127 Pisa, Italy}

\author{A. Singha}
\affiliation{NEST, Istituto Nanoscienze-CNR and Scuola Normale Superiore, I-56127 Pisa, Italy}

\author{M. Gibertini}
\affiliation{NEST, Istituto Nanoscienze-CNR and Scuola Normale Superiore, I-56127 Pisa, Italy}

\author{B. Karmakar}
\affiliation{NEST, Istituto Nanoscienze-CNR and Scuola Normale Superiore, I-56127 Pisa, Italy}

\author{M. Polini}
\affiliation{NEST, Istituto Nanoscienze-CNR and Scuola Normale Superiore, I-56127 Pisa, Italy}

\author{V. Piazza}
\affiliation{NEST, Istituto Nanoscienze-CNR and Scuola Normale Superiore, I-56127 Pisa, Italy}

\author{L.N. Pfeiffer} 
\affiliation{Department of Electrical Engineering, Princeton University, NJ USA}

\author{K.W. West}
\affiliation{Department of Electrical Engineering, Princeton University, NJ USA}

\author{F. Beltram}
\affiliation{NEST, Istituto Nanoscienze-CNR and Scuola Normale Superiore, I-56127 Pisa, Italy}

\author{V. Pellegrini}
\affiliation{NEST, Istituto Nanoscienze-CNR and Scuola Normale Superiore, I-56127 Pisa, Italy}

\date{\today}   

\begin{abstract}
We report the magneto-transport properties of a two-dimensional electron gas in a modulation-doped AlGaAs/GaAs heterostructure subjected to a lateral potential with honeycomb geometry. Periodic oscillations of the magneto-resistance and a delocalized-localized transition are shown by applying a gate voltage. We argue that electrons in such artificial-graphene lattices offer a promising approach for the simulation of quantum phases dictated by Coulomb interactions.
\end{abstract}


\maketitle 
														
Magneto-transport properties of two-dimensional electron systems confined in semiconductor heterostructures and subjected to quantizing 
in-plane potentials were extensively investigated~\cite{albrecht,melinte,weiss,weiss1,paris}. 
Recently renewed interest on these systems was triggered by theoretical proposals for the realization of quantum simulators in scalable solid-state 
systems~\cite{yama1,yama2,nori}. These authors discussed the evaluation of Coulomb interaction strength and hopping properties of electrons confined in square arrays of coupled GaAs quantum dots or at the minima of standing waves created by surface acoustic waves. Such quantum simulators can represent a viable strategy to investigate quantum phases and many-body effects in a controlled way \cite{feynman,loyd} and may open the way to the emulation of quantum phenomena not easily accessible in nature. To date, however, experimental tests of quantum simulators were carried out only in few systems including in particular cold atoms~\cite{cold1,cold2} and trapped ions~\cite{ion}. 
\par
Here we report on the fabrication and magneto-transport characterization of a scalable quantum simulator obtained in a modulation-doped high-mobility GaAs semiconductor heterostructure. The structure consists of an array of pillars with honeycomb symmetry that are formed on the surface of the semiconductor by shallow etching. Owing to the dependence of band-bending profiles on GaAs cap-layer thickness~\cite{leonid}, these nanofabricated pillars induce a lateral potential modulation ($V_0$) of amplitude of few meV's acting on the electronic system. This lateral potential manifests itself by the appearance of a modulation of the magneto-resistivity that is periodic in the magnetic field. 
\par
We shall show that a transition from a conducting to an insulating regime can be induced by changing the voltage of a metallic gate deposited on top of the array. Our experimental findings suggest that while electrons occupy delocalized states at low values of the gate voltage ({\it i.e.} when carrier density is high and $V_0$ is small), they localize in the confined levels of the potential minima underneath the pillars as we sweep the gate voltage and increase $|V_0|$. This interpretation is supported by theoretical evaluations of the hopping parameters for these quantum-confined states. 
\begin{figure}[!ht]
	\centering
	\includegraphics*[width=8.5cm]{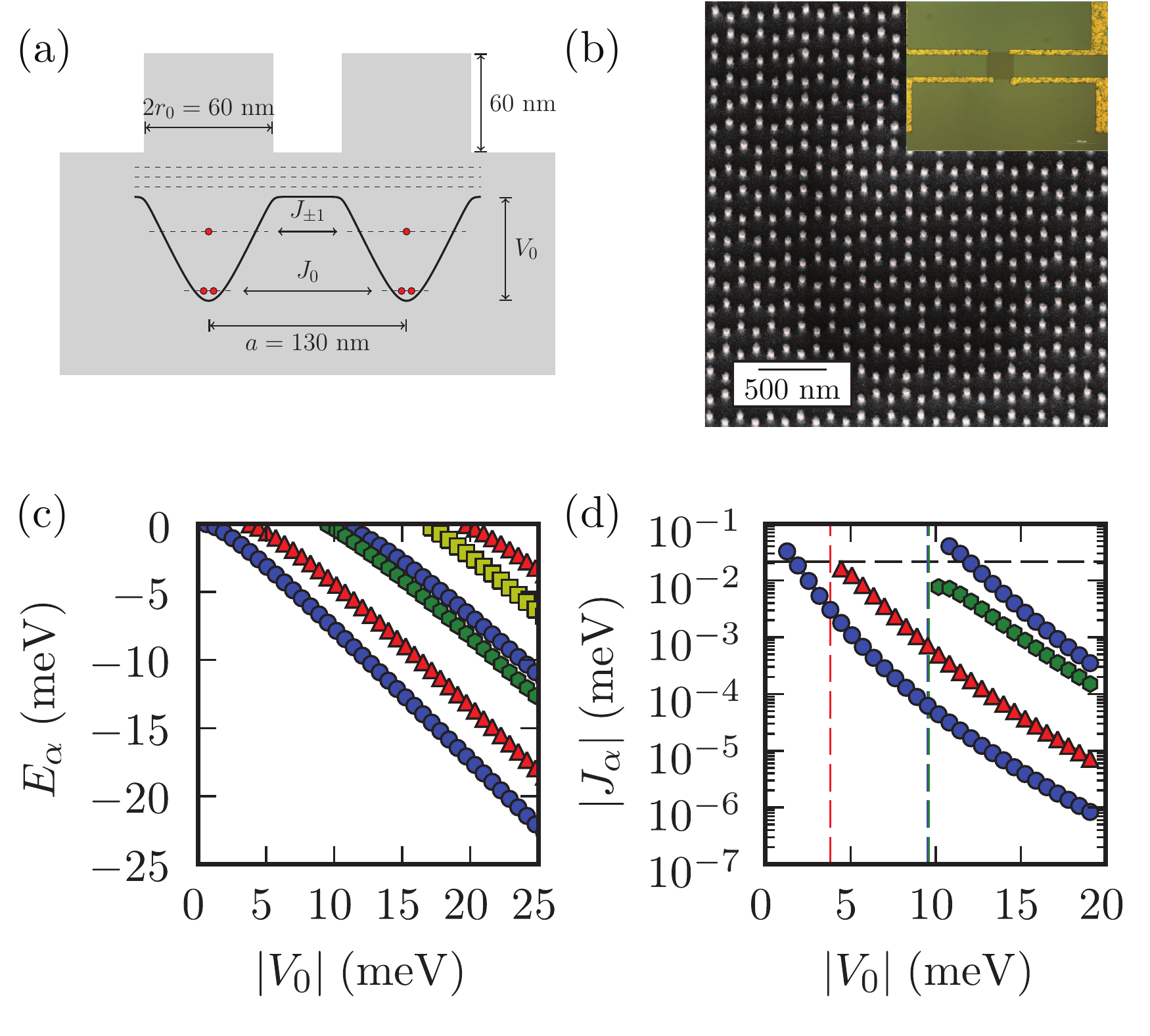}
	\caption{(a) Sketch of the lateral potential with relevant geometrical parameters. $J_\alpha$ denotes the hopping 
    parameter and $\alpha = \{n,m\}$ a set of quantum numbers (see text). (b) Scanning electron microscopy image of the nanopatterned modulation-doped GaAs/AlGaAs sample with  honeycomb geometry. 
	The inset shows a picture of the device with four Ohmic contacts. (c) Calculated bound-state energy levels $E_\alpha$  as functions of $|V_0|$. Different curves correspond to different values of the angular 
      momentum $m$: $m =0$ (blue circles), $m=\pm 1$ (red triangles), $m=\pm 2$ (green hexagons), and $m = \pm 3$ (yellow squares). Curves corresponding to $m$ values which differ only by a sign are degenerate. 
For a fixed value of $V_0$ curves with same symbol and color coding but higher energy correspond to excited states with different radial quantum numbers.
(d) The 
    hopping integrals $J_\alpha$ as functions of $|V_0|$. Symbol coding is the same as in panel (c). The horizontal dashed line 
	labels the thermal energy at $T=250~{\rm mK}$. Vertical dashed lines mark the threshold values of $V_0$ at which a new bound state 
      appears. The color of the dashed lines is related to the value of $m$ as in panel (c). \label{fig:one}}
\end{figure}
\par
The demonstration that delocalized and localized electron states can be realized and controlled in these artificial honeycomb lattices with the same topology of graphene~\cite{gibertini} opens up exciting opportunities for the study of unusual electronic phases and collective modes induced by the lattice potential at ultra-low temperatures and for the simulation of massless Dirac fermion~\cite{gibertini} ground states in the presence of external fields. Of particular relevance is the possibility to explore a regime of strong correlations that is not easily achievable in graphene and is characterized by inter-pillar hopping being much smaller than the on-site Coulomb energy.
\par
The sample used in this study contains a two-dimensional electron gas (2DEG) in a $25~{\rm nm}$ wide, one-side modulation-doped 
Al$_{0.1}$Ga$_{0.9}$As/GaAs  quantum well (QW). The 2DEG is positioned $170~{\rm nm}$ underneath the surface 
(the doping layer is at $110~{\rm nm}$). It has low-temperature electron density $n_{\rm e} = 1.1 \times 10^{11}~{\rm cm}^{-2}$ and mobility 
of $2.7 \times 10^6~{\rm cm}^{2}/({\rm V s})$. The in-plane potential modulation was achieved by defining an array of Nickel disks (with 
diameter $2r_0$) arranged in a honeycomb-lattice geometry (with lattice constant $a$) by e-beam nanolithography and then by etching away the 
material outside the disks by inductive coupled reactive ion shallow etching~\cite{garcia,gibertini}. Different etching depths $d$ were realized. 
Results presented here were obtained with $d = 60~{\rm nm}$ [see Fig.~\ref{fig:one}(a)]. Larger etching depth lead to electron localization even 
at zero gate voltage. The resulting lateral nano-patterning after removal of the Nickel, which extends on a $100~{\rm \mu m}\times 100~{\rm \mu m}$ 
square region, is shown in Fig.~\ref{fig:one}(b). Experimental values of the parameters are $a \sim 130~{\rm nm}$ and $r_0 \sim 30~{\rm nm}$. 
\begin{figure}[!ht]
	\centering
	\includegraphics*[width=8.5cm]{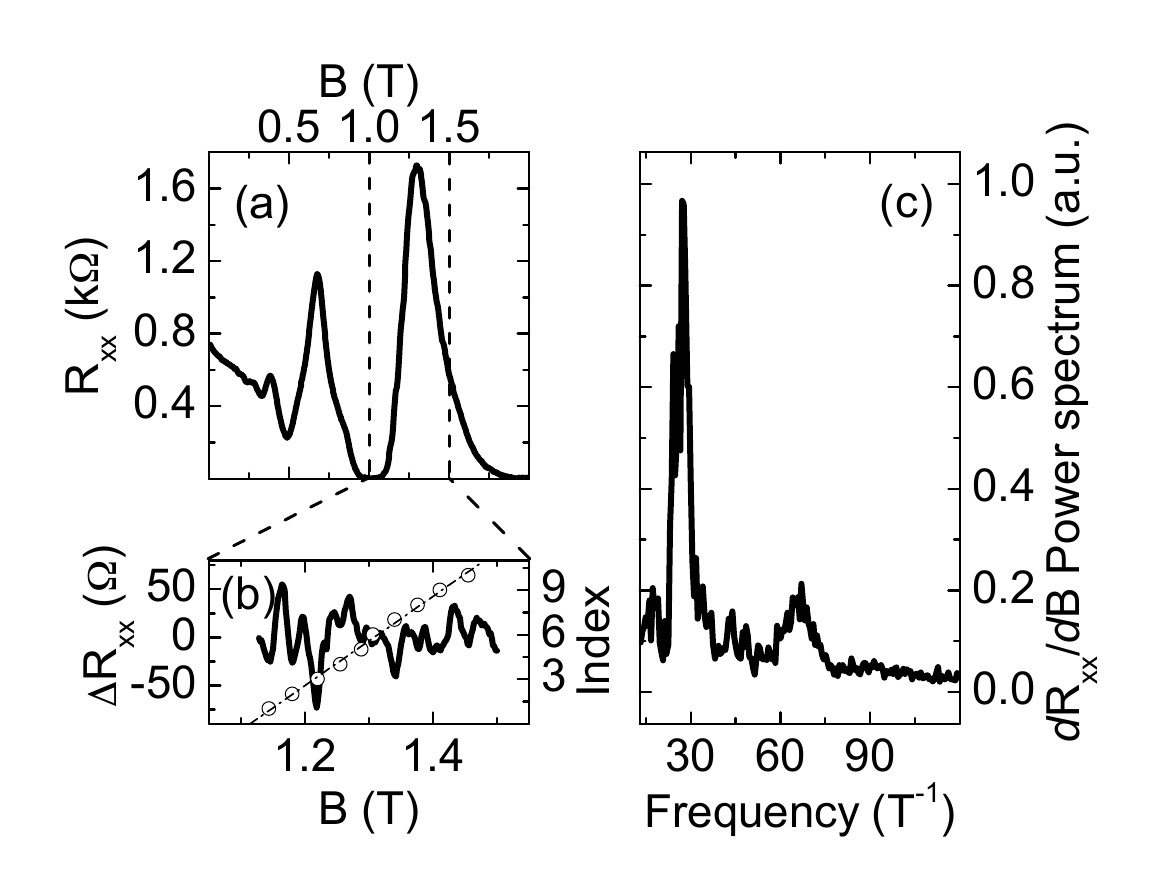}
	\caption{(a) Longitudinal resistance $R_{xx}$ as a function of the magnetic field at a gate voltage $V_{\rm g} = -1.62~{\rm V}$ 
	and charge density $n = 4.6 \times 10^{10}~{\rm cm}^{-2}$. 
	(b)  A magnified portion of data between $1.1~{\rm  T}$ and $1.5~{\rm  T}$. A polynomial fit of the main $R_{xx}$ curve was subtracted 
	to enhance the visibility of  magnetic-field-periodic oscillation. A scatter-plot of the resistance minima position is also shown.  
	(c) Device oscillation power spectrum obtained by applying the Fast Fourier Transform algorithm on a set of $dR_{xx}/dB$ taken 
	at several gate voltages (from $-1.2$ to $-1.62~{\rm V}$, corresponding to $n=4.8$ and $4.5 \times 10^{10}~{\rm cm}^{-2}$, respectively).
	Contributions from all curves have been averaged to reduce the spectral noise.\label{fig:two}}
\end{figure}
\par
In order to probe the transport properties of the device, Ni/AuGe/Ni/Au (10/180/10/100 nm) annealed ohmic contacts were evaporated at 
the corners of the array [see inset of Fig.~\ref{fig:one}(b)]. Other contacts were realized off the array where the $60~{\rm nm}$-thick 
GaAs layer was uniformly etched away. Thanks to the low level of damage introduced by this process~\cite{garcia}, the 60nm-thick pillars of 
the honeycomb array act as a {\it weak} attractive potential ($V_0 <0$) for electrons. The absence of any signature of conductive behavior 
outside the array (data not shown) suggests that when the etching affects the whole sample surface it is sufficient to fully deplete the QW. We 
therefore estimate $V_0 \approx E_{\rm F} \approx 4~{\rm meV}$ where $E_{\rm F}$ is the Fermi energy of the original unperturbed 2DEG. 
A $120~{\rm nm}$-thick Al gate was deposited on top of the device to allow the tuning of the charge density and $V_0$. Electrical measurements were performed by a lock-in technique in a filtered $^3$He magneto-cryostat with 250 mK base temperature. While the overall behavior was the same after different cool-downs with different illumination procedures, the values of the charge density at a given gate voltage were found to change from cool-down to cool-down.   
Figure~\ref{fig:two}(a) shows a representative low-temperature longitudinal magneto-resistance characteristic measured at gate 
voltage $V_{\rm g} = -1.62~{\rm V}$. The observation of distinct quantum Hall signatures  with an estimated electron 
density of $4.5 \times 10^{10}~{\rm cm}^{-2}$ demonstrates the existence of delocalized states populated by electrons 
at this density. Weak modulations of the resistivity [highlighted in Fig.~\ref{fig:two}(b)] in the magnetic-field range between filling 
factors $\nu =2$ and $ \nu =1 $ are present and appear only below a certain gate voltage. In Fig.~\ref{fig:two}(c) we report the averaged 
Fast Fourier Transform (FFT) relative to a set of $dR_{xx}/dB$ traces taken from $V_{\rm g} = -1.2~{\rm V}$  to $-1.62~{\rm V}$.  Two main 
peaks are observed at 27 T$^{-1}$ (FWHM 6 T$^{-1}$) and 67 T$^{-1}$ (FWHM 20 $T^{-1}$) corresponding, respectively, to 2.5 and 6.5 times the 
geometrical area of the hexagonal unit cell. Following early works based on a semiclassical analysis~\cite{weiss1,fleisch}, the FFT 
peaks can be attributed to the quantization of periodic orbits, although a precise identification requires a more extended theoretical 
investigation.
\begin{figure}[!hb]
	\centering
	\includegraphics*[width=8.5cm]{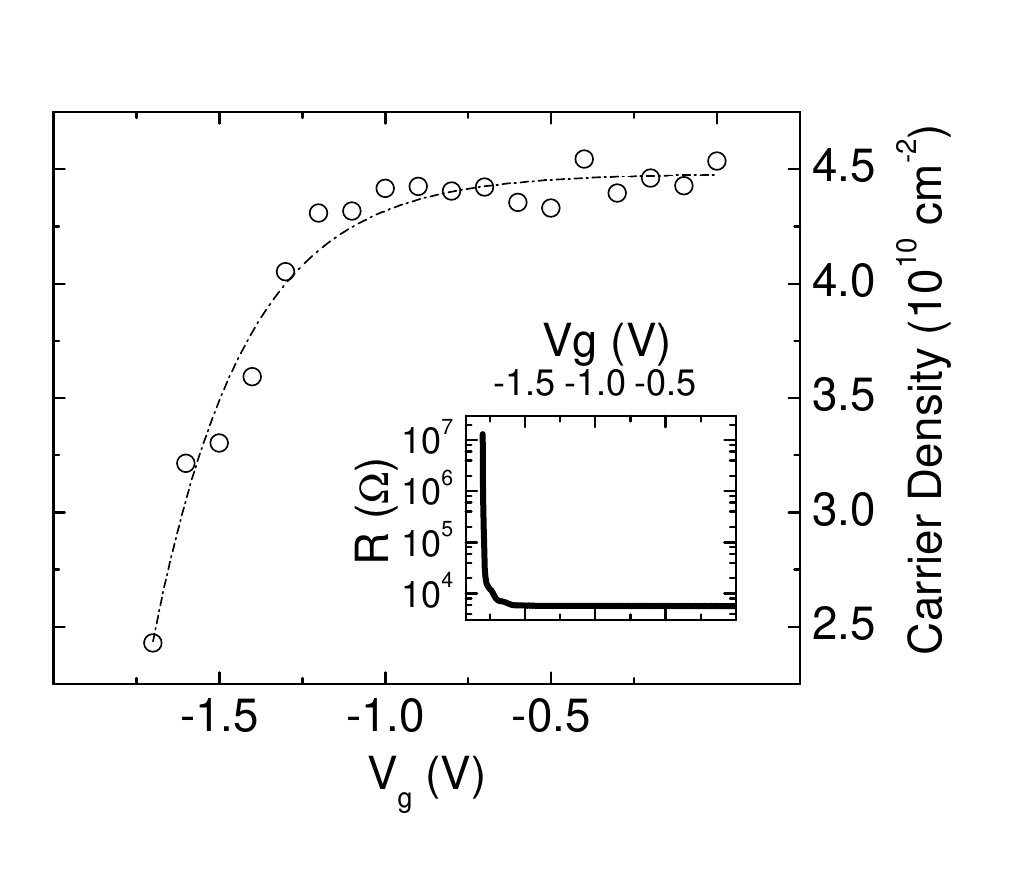}
	\caption{Carrier density (after a representative cooldown) as a function of gate voltage 
	$V_{\rm g}$ at $T= 250~{\rm mK}$. The inset shows the behavior of the longitudinal resistance measured {\it via} 
	a two-terminal setup as a function of $V_{\rm g}$ at 250~${\rm mK}$ and zero magnetic field. 
	The localized regime is achieved below $V_{\rm g} = -1.5~{\rm V}$. \label{fig:three}}
\end{figure} 
\par
We now address the evolution of the carrier density as a function of $V_{\rm g}$ in a set of two-wire quantum Hall measurements. The results 
presented in Fig.~\ref{fig:three} display the expected decrease of the charge density above a threshold gate voltage whose value is associated with 
the depletion of the residual charge trapped in the Si doping layer.  We remark that as we lower $V_{\rm g}$ the resistance becomes so high that it hinders a reliable estimate of the electron density. This is consistent with a transition to a regime of electrons localized in the minima of the honeycomb potential as supported by the finite density of $2.5-3 \times 10^{10}~{\rm cm}^{-2}$ found in the transition region at $V_{\rm g} \approx -1.5~{\rm V}$ . We remark that the two-wire configuration setup allows us to estimate the charge density even in this transition region. This is otherwise impossible because of the non-linear behavior of the voltage-probe contacts in that regime. 
\par
In order to investigate in more detail the transition to the localized-insulating regime we present in the inset to Fig.~\ref{fig:three} the
sample resistance as a function of $V_{\rm g}$ at zero perpendicular magnetic field. A transition associated to an abrupt increase of the sample resistance leads to a persistent insulating state. While the transition threshold was observed to depend weakly ($\pm 200~{\rm mV}$) on the cool-down/illumination procedure, the qualitative behavior remains the same from cool-down to cool-down. The sample can be driven again to the conductive/delocalized regime by means of a thermal cycle ($250~{\rm mK}-30~{\rm K}-250~{\rm mK}$) or by illumination. This latter behavior can be related to the presence of several trapping mechanisms largely due to the device fabrication steps.
\par 
The insulating behavior highlighted in Fig.~\ref{fig:three} can be explained in terms of electron localization 
in the confined levels of the potential minima as shown in 
Fig.~\ref{fig:one}(a). To confirm this picture, which requires negligible inter-site hopping between the confined levels of neighboring minima, we evaluated the hopping integrals between states in two pillars placed at a relative center-to-center distance equal to $a$. To this end, the external potential created by a single pillar was modeled by a simple rotationally-invariant piecewise-constant function of the 2D position ${\bm r}$: $V_{\rm ext}(r) = V_0$ if $r = |{\bm r}| < r_0$ and zero otherwise. The resulting eigenfunctions are particularly simple 
in polar coordinates $(r,\theta)$, where they can be written as~\cite{quantumnumbers} $\Phi_\alpha({\bm r}) = \psi_{n,m}(r) \chi_m(\theta)$, $\chi_m(\theta) = e^{i m\theta}/\sqrt{2 \pi}$, with $m$ a relative integer, being the eigenstates of the angular momentum operator ${\hat L}_z = -i\hbar \partial/\partial \theta$ along the ${\hat {\bm z}}$ direction. 

The spectrum of bound states with energy $E_\alpha<0$ can be found by solving the following transcendental equation: 
\begin{equation}\label{eq:consistBzero}
k~\frac{J'_m(k r_0)}{J_m(k r_0)} = \kappa~\frac{K'_m(\kappa r_0)}{K_m(\kappa r_0)}~.
\end{equation}
Here $k = [2 m_{\rm b} (|V_0|-|E_\alpha|)/\hbar^2]^{1/2}$, $\kappa = (2 m_{\rm b}|E_\alpha|/\hbar^2)^{1/2}$, $m_{\rm b}$ is the band mass, and $J_m(x), K_m(x)$ are standard Bessel functions [$J'_m(x)$ and $K'_m(x)$ begin their derivatives with respect to $x$]. The spectrum $E_\alpha$ obtained by solving Eq.~(\ref{eq:consistBzero}) is reported in Fig.~\ref{fig:one}(c) as a function of $V_0$ for various values 
of the angular momentum $m$. 
\par
The hopping parameter $J_\alpha$, which provides the rate at which tunneling occurs from a pillar to a neighboring one, can be calculated from the following two-center integral
\begin{equation}\label{eq:hopping}
J_\alpha = \int d^2 {\bm r}~\Phi^*_\alpha({\bm r}) V_{\rm ext}(r) \Phi_\alpha({\bm r} - {\bm a})~.
\end{equation}
Here $\Phi_\alpha({\bm r})$ are the eigenstates of a single pillar centered at the origin and ${\bm a}$ is a nearest-neighbor position vector.  
Numerical results for $J_\alpha$ are reported in Fig.~\ref{fig:one}(d). We see that in the relevant range of potential depths, $V_0 \approx 5~{\rm meV}$, the hopping integral 
associated with the two lowest bound state is smaller than the thermal energy (horizontal dashed line).
In the presence of a magnetic field ${\bm B} = B {\hat {\bm z}}$ it is easy to show {\it via} a gauge transformation that
\begin{equation}\label{eq:finalJalpha}
J_\alpha = \int d^2 {\bm r}~e^{2\pi i\frac{\varphi({\bm r})}{\varphi_0}}~\Phi^*_\alpha({\bm r}) V_{\rm ext}(r) \Phi_\alpha({\bm r}-{\bm a})~,
\end{equation}
where we have defined $\varphi({\bm r}) = {\bm B} \cdot ({\bm r} \times {\bm a})/2$ and $\varphi_0 = \hbar c/e$ is the flux quantum. 
The complex phase factor in Eq.~(\ref{eq:finalJalpha}), which resembles the Peierls phase, and the magnetic-field dependence of the 
eigenstates $\Phi_\alpha({\bm r})$ lead to a further (exponential) suppression of $J_\alpha$ (data not shown). 
\par
In conclusion, we have studied electron transport in a high-mobility GaAs quantum well in which electrons are subjected to a lateral 
artificial potential with honeycomb geometry. The possibility to control the localized/delocalized nature of the electronic states suggests 
further use of this system for quantum simulations in a regime dominated by electron correlations. 

{\it Acknowledgements.}--- We thank A. Pinczuk for useful discussions.


\begin{thebibliography}{40}
\bibitem{albrecht} C. Albrecht {\it et al.}, Phys. Rev. Lett. {\bf 86}, 147 (2001).
\bibitem{melinte} S. Melinte {\it et al.}, Phys. Rev. Lett. {\bf 92}, 036802 (2004).
\bibitem{weiss} R. R. Gerhardts {\it et al.}, Phys. Rev. Lett. {\bf 62}, 1173 (1989).
\bibitem{weiss1} D. Weiss {\it et al.}, Phys. Rev. Lett. {\bf 70}, 4118 (1993).
\bibitem{paris} E. Paris {\it et al.}, J. Phys.: Condens. Matter {\bf 3}, 6605 (1991).
\bibitem{yama1} T. Byrnes {\it et al.}, Phys. Rev. Lett. {\bf 99}, 016405 (2007).
\bibitem{yama2} T. Byrnes {\it et al.}, Phys. Rev. B {\bf 78}, 075320 (2008).
\bibitem{nori} I. Buluta and F. Nori, Science {\bf 326}, 108 (2009).
\bibitem{feynman} R. P. Feynman, Int. J. Theor. Phys. {\bf 21}, 467 (1982). 
\bibitem{loyd} S. Lloyd, Science {\bf 274}, 1073 (1996).
\bibitem{cold1} M. Greiner {\it et al.}, Nature {\bf 415}, 39 (2002).
\bibitem{cold2} G. B. Jo {\it et al.}, Science {\bf 325}, 1521 (2009).
\bibitem{ion} A. Friedenauer {\it et al.}, Nature Phys. {\bf 4}, 757 (2008). 
\bibitem{leonid} L. P. Rokhinson {\it et al.}, Superlattices and Microstructures, {\bf 32}, 99 (2002).
\bibitem{gibertini} M. Gibertini {\it et al.}, Phys. Rev. B {\bf 79}, 241406(R) (2009).
\bibitem{garcia} C. P. Garcia {\it et al.}, Phys. Rev. Lett. {\bf 95}, 266806 (2005).
\bibitem{fleisch} R. Fleischmann, T. Geisel, and R. Ketzmerick, Phys. Rev. Lett. {\bf 68}, 1367 (1992).
\bibitem{quantumnumbers} $\alpha =\{n,m\}$ denotes a set of quantum numbers comprising 
the radial quantum number $n$ (which sets the zeroes of the radial wavefunction $\psi_{n,m}(r)$ 
and the angular momentum quantum number $m$.

\end{thebibliography}
\end{document}